\documentclass{pasj01}

\begin{document}

\author{Mariko \textsc{Kimura}\altaffilmark{1,*}
        and Yoji \textsc{Osaki}\altaffilmark{2}
        }
\email{mariko.kimura@riken.jp}

\altaffiltext{1}{Institute of Physical and Chemical Research (RIKEN), 2-1 Hirosawa,
Wako, Saitama 351-0198}
\altaffiltext{2}{Department of Astronomy, School of Science, University of Tokyo, Hongo, Tokyo 113-0033}

\title{
KIC 9406652: A laboratory for tilted disks in cataclysmic variable stars.~II.~Modeling of the orbital light curves
}

\Received{} \Accepted{}


\SetRunningHead{Kimura and Osaki.}{Modeling of the orbital light curves of KIC 9406652}

\maketitle

\begin{abstract}

KIC 9406652, one of the recently identified IW And-type 
dwarf novae, is the best target for studying the tilted disk 
in cataclysmic variable stars.  
In a previous paper by Kimura, Osaki, and Kato (2020),
we analyzed its {\it Kepler} light curves and found that 
its orbital light curves during the brightening stage 
were dominated by the reflection effect of the secondary star 
and varied with the orientation of the tilted disk;
the amplitude was maximized at the minimum of the super-orbital 
signal and the phase of the light maximum shifted to 
an earlier one with the advance of the super-orbital phase.  
We argued there that this was the direct evidence of 
the retrogradely precessing tilted disk as the secondary star 
acts like a reflecting object.  
In order to confirm this interpretation, we have performed 
numerical modeling of orbital light curves in this paper.  
We have succeeded in reproducing the main characteristics of 
the observed orbital light curves by a simple model in which 
the secondary star is irradiated by the tilted disk.  
We have also constrained the inclination angle, $i$, of 
the binary system and the tilt angle, $\theta$, of the disk 
purely from photometric considerations.  
The best-fitting parameter set is found to be $i \sim$45~deg 
and $\theta \sim$2.0~deg, respectively.  
The orbital inclination thus estimated is consistent with 
that obtained from the spectroscopic considerations within 
the uncertainty limit.  
On the other hand, the tilt angle of the disk could be 
underestimated by using only the semi-amplitude of 
super-orbital signals.  

\end{abstract}

\section{Introduction}

In the previous paper (\cite{kim20kic940}; hereafter 
referred to as Paper I), we have extensively studied 
the {\it Kepler} light curve of a cataclysmic variable star, 
KIC 9406652.  There we have demonstrated that this star is 
one of the best laboratories for tilted disks in cataclysmic 
variables (CVs) as it offers the first direct evidence that 
the accretion disk is in fact tilted out of the binary orbital 
plane and retrogradely precessing from analyses of light 
curves of three periodic signals: orbital, negative superhump, 
and super-orbital signals.  The star, KIC 9406652, is 
a CV sub-classified as IW And stars, or anomalous Z Cam stars.  
IW And stars are a small group of dwarf novae (DNe) which 
exhibit a characteristic light curve: a repetition of 
a quasi-standstill (i.e., a mid-brightness interval with 
(damping) oscillations) terminated by small-amplitude 
brightening, which is called ``IW And-type phenomenon'' 
by \citet{kat19iwand}.  The typical repetition cycle is 
around 50 days.  More detailed information on IW And stars 
is given in the introduction of Paper I.  
The present paper is an extension of Paper I, in which 
we try to model the orbital light curves in this star 
based on the tilted disk.  

The signature of the tilted disk manifests itself in 
the light curve most clearly by the two periodic light 
variations, i.e., negative superhumps and 
super-orbital signals.  
Negative superhumps are small-amplitude periodic light 
modulations having a period slightly shorter than the 
orbital period, and super-orbital signals are wavy light 
modulations with a period much longer than the orbital period 
(typically around 20 times of the orbital one).  
\textcolor{black}{
In the research on CVs, these two periodic signals 
were discovered for the first time in TV Col 
\citep{mot81tvcol,bon85tvcol} and \citet{bon85tvcol} 
interpreted that these two periodic signals originate 
from a retrogradely precessing tilted accretion disk, 
inspired by the observations of some X-ray binaries 
(e.g., \cite{kat73hzher,lei84v1343aql}).  
Negative superhumps and/or super-orbital signals in 
various types of CVs were detected in the subsequent 
observational studies 
(e.g., \cite{har95v503cyg,pat97v603aql,arm13aqmenimeri}), and 
the source of these periodicities has been investigated.  
Currently, negative superhumps are thought to be produced by 
variable brightness of the bright spot sweeping on 
the tilted disk surface \citep{woo00SH,mur02warpeddisk,woo07negSH}.  
In line with this interpretation, super-orbital signals are 
thought to be produced by varying projection area of 
the tilted disk to the observer as its nodal line slowly 
precesses \citep{pat99SH,pat02v442ophj1643}.  
Besides, various mechanisms for the origin of disk tilt 
have been proposed, but no consensus has been reached.  
For instance, \citet{lub92tilt} suggested that the 3:1 
resonance may amplify perturbations of a disk in 
the vertical direction.  
\citet{sma09negativeSH} advocated a model in that 
variable vertical velocity component of the gas stream, 
which arises from variable irradiation of the secondary 
star by the primary star due to the shadow of a tilted, 
precessing disk, maintains disk tilt.  
\citet{mon10disktilt} proposed that the disk is forced 
to be tilted by a kind of buoyancy called ``lift'' and 
this idea was tested by SPH simulations 
\citep{mon12negSHSPH}.}  

KIC 9406652, is an ideal star to study the tilted disk 
because it exhibits both negative superhumps and 
super-orbital light modulations and because 
the comprehensive {\it Kepler} light curve extending 
over 4 years is available \citep{gie13j1922}.  
We analyzed its {\it Kepler} light curve in Paper I and 
one of the most important findings concerns the orbital 
light curves in the brightening stage of the IW And-type 
phenomenon.  
It has turned out that the orbital light curve is 
dominated by the effect of irradiation of the secondary star 
in the brightening stage (see the top panel of Fig.~8 of 
Paper I) and the orbital light curve varied greatly with 
the super-orbital phase, i.e., with the orientation of 
the tilted disk to the observer.  
The amplitude of the orbital light curve was found to be 
maximized at the minimum of the super-orbital signals, 
i.e., when the orientation of the tilted disk is directed 
opposite to the observer, and the maximum of the orbital 
light curve was found to be shifted to earlier orbital phases 
as the super-orbital phase advances (see, Fig.~14 of Paper I).  
We argued in Paper I that the above-mentioned fact 
offers direct evidence that the accretion disk is tilted 
out of the binary orbital plane and its nodal line rotates 
retrogradely as the secondary star acts like a reflecting 
mirror for the tilted disk.  

In Paper I, we did not subtract negative superhump 
signals from the original light curve in constructing 
the orbital light curves. 
However, since the subtraction of negative superhump 
signals affects the orbital light curves greatly, 
we \textcolor{black}{first} perform this in Sec~2.  
In Sec.~3, we describe the basic model and the numerical 
code that we use for the light curve modeling.  
In Sec.~4, we set the parameters used for modeling.  
Sec.~5 presents the results of modeling. 
Sec.~6 is the summary and discussion.

\section{Subtraction of negative superhump signals from orbital light curves}

In this study, we try to model the orbital light curves 
during the brightening stage based on a model of 
the tilted disk and the irradiated secondary star.  
To do so, we must first mention that some revisions are 
needed to the top panel of Fig.~8 and Fig.~14 of Paper I.  
In constructing the orbital light curves in these figures, 
we did not subtract the effect of negative superhumps.  
Since the beat period between the orbital period and 
the negative superhump period is exactly equal to 
the super-orbital period, the orbital light curves 
with the super-orbital phase are strongly affected by 
contamination of negative superhumps and it is essential to 
subtract this effect.  

\begin{figure}[htb]
\begin{center}
\FigureFile(80mm, 50mm){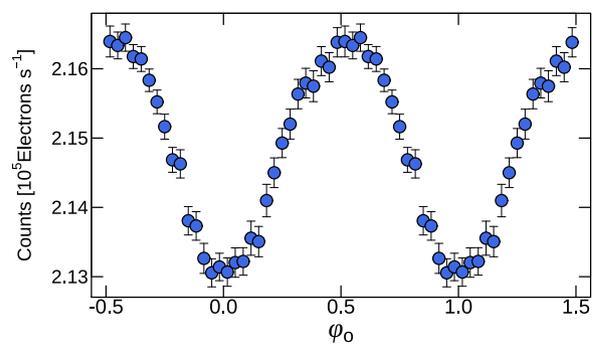}
\end{center}
\caption{Phase-averaged profile of orbital signals during brightening in KIC 9406652, which has been constructed after subtracting negative superhump signals from the original data set.  
}
\label{orb-b}
\end{figure}

\begin{figure*}[htb]
\begin{center}
\setlength{\fboxrule}{0.7pt}
\begin{minipage}{0.49\hsize}
\fbox{
\FigureFile(80mm, 50mm){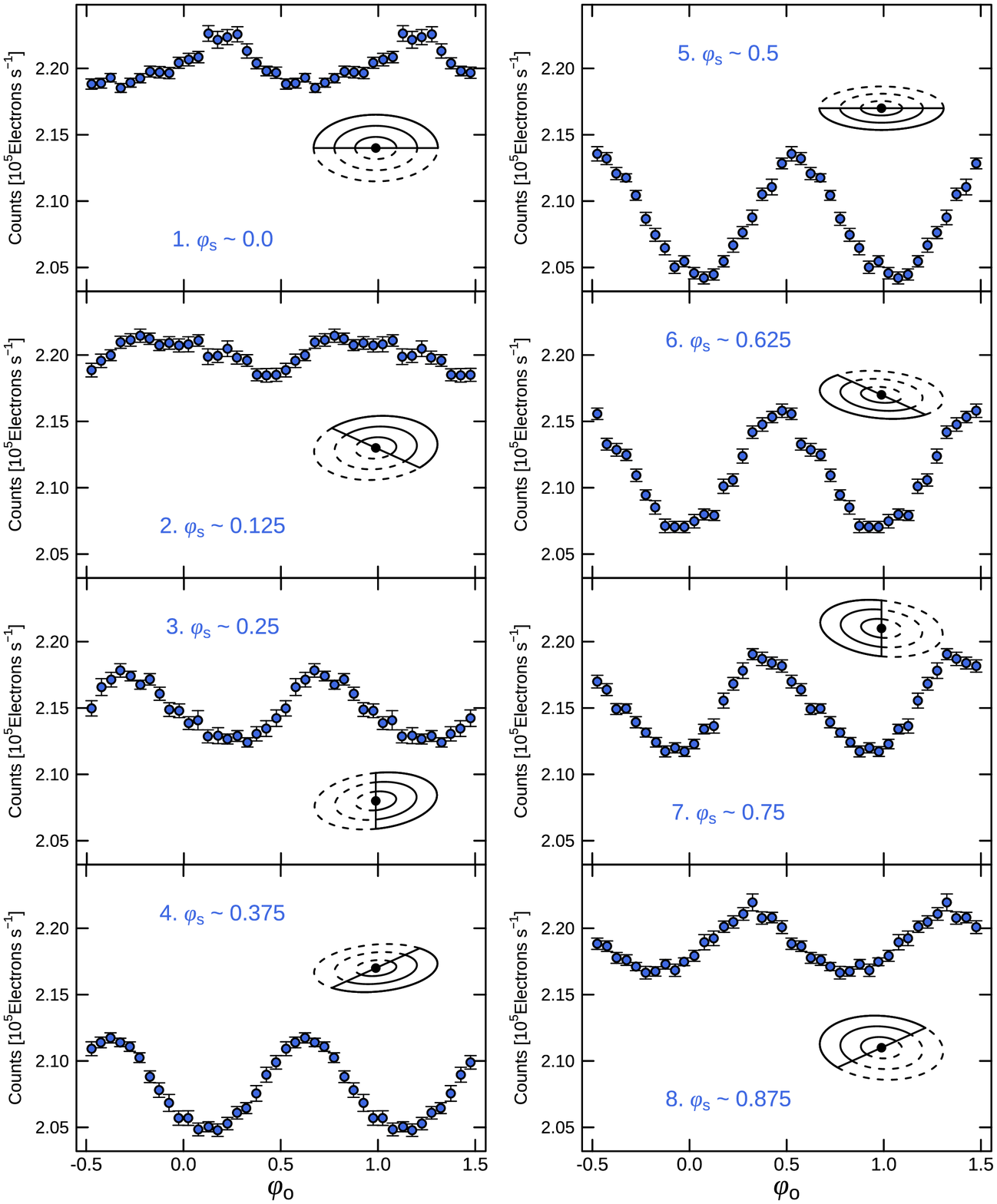}
}
\end{minipage}
\begin{minipage}{0.49\hsize}
\fbox{
\FigureFile(80mm, 50mm){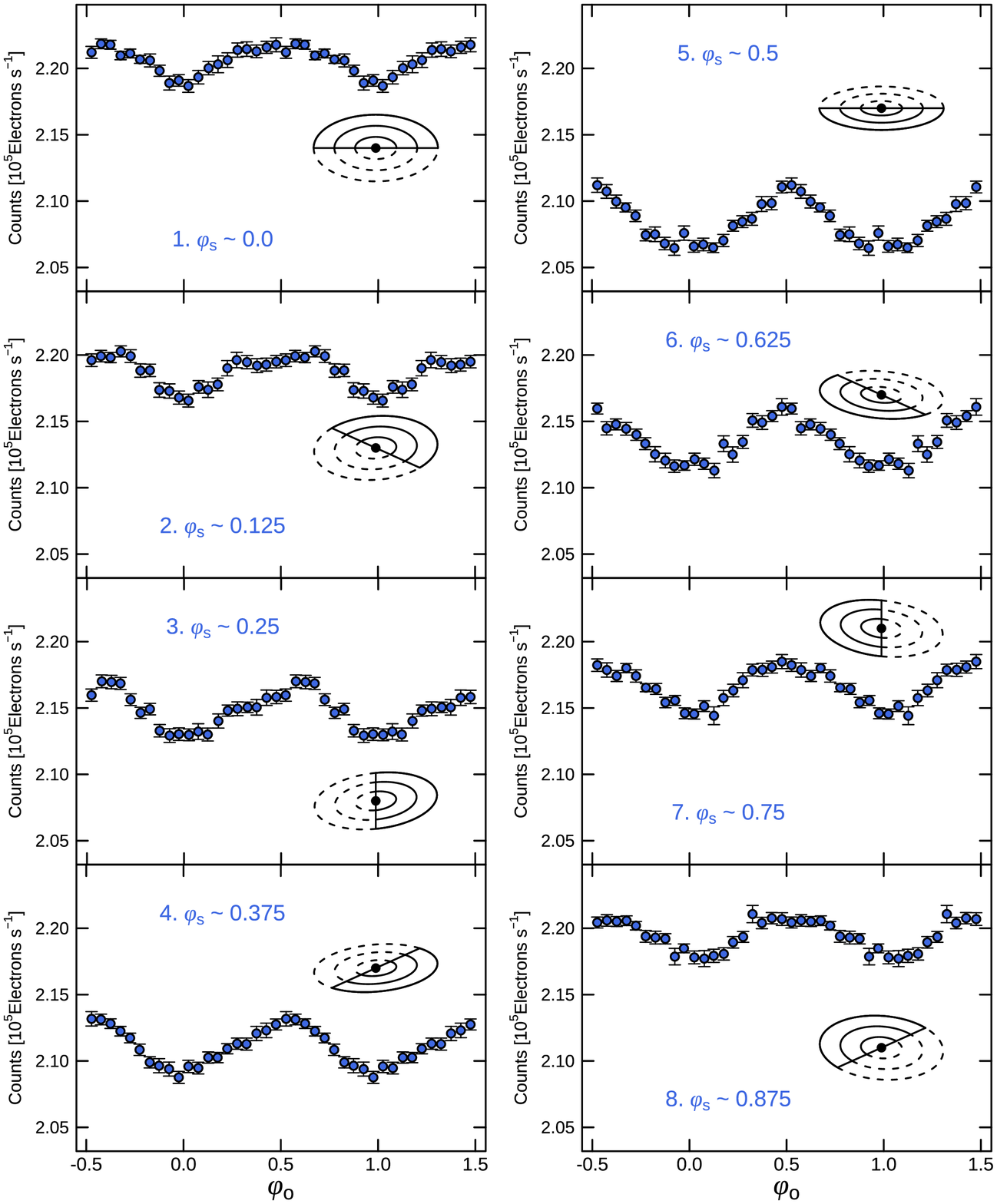}
}
\end{minipage}
\end{center}
\caption{Comparison of orbital light curves before and after the subtraction of 
negative superhump signals for eight types of orbital profiles from the data limited for brightening in KIC 9406652. The left figure is the original version same as that in Fig.~14 of Paper I and the right figure is the revised version after the subtraction of negative superhumps.  The inset of each figure shows the orientation of the tilted disk where the central WD is marked by the black dot and the diametric line passing the WD is the nodal line of the tilted disk.  The solid and dashed lines represent the equatorial plane of the tilted disk above and below the orbital plane of the binary system, respectively.  In the right figure, the mean flux level of each light curve is calculated by using the orange circles in Figure \ref{sup-b}.  
}
\label{orb8b}
\end{figure*}

We here subtract negative superhump signals from 
the original light curve, and reconstruct the orbital 
light curves.  
Since the period and light curve profile of negative 
superhumps were time-varying and the profile sometimes 
depended on the luminosity of KIC 9406652 (see Fig.~1, 
Fig.~2, and the left panel of Fig.~3 in Paper I), 
we divide the observational data into shorter intervals 
in which we may regard that they are more or less  
constant.  We then construct an average light curve 
profile of negative superhumps for each interval and 
subtract it from the original light curve.  
We finally construct the phase-averaged orbital light curves 
from the processed light curve in the brightening stage 
with the same manner as described in subsections 3.3 and 3.5 
of Paper I.  

Figure \ref{orb-b} and the right panel of Figure \ref{orb8b} 
are the revised versions to the top panel of Fig.~8 and 
Fig.~14 of Paper I.  For comparison, we also show 
the original one (Fig.~14 of Paper I) in the left panel 
of Figure \ref{orb8b}.  
Figure \ref{orb-b} shows the orbital light curve for all data 
during the brightening stage where $\varphi_{\rm o}$ is 
the orbital phase and its phase zero is defined by 
the inferior conjunction of the secondary star.  
Figure \ref{orb8b} presents eight orbital light curves 
for eight different super-orbital phases $\varphi_{\rm s}$ 
during the brightening phase.  
Here the phase zero is defined at the maximum light of 
super-orbital light modulations and it corresponds to 
the phase when the front side of the tilted disk turns 
toward the observer.  

Figure \ref{orb-b} is almost the same as the original 
one of Fig.~8 of Paper I, 
which means that the orbital profile averaged for all 
super-orbital phases does not suffer very much from 
negative superhumps.  
On the other hand, the orbital light curves for specific 
super-orbital phases are modified very much by the subtraction 
of negative superhumps (compare the left and right panels 
of Figure \ref{orb8b}).  
After the subtraction of negative superhumps, we see that 
(1) the difference in the amplitude becomes smaller 
between the orbital light curve at $\varphi_{\rm s} \sim 0.0$ 
and that at $\varphi_{\rm s} \sim 0.5$, (2) the phase shift 
of the light maximum of orbital light curves with the advance 
of super-orbital phases becomes less conspicuous, and 
(3) the light minima of orbital light curves always occur 
at $\varphi_{\rm o} = 0.0$ while it was not the case 
in the original ones.  
However, the main characteristics remain the same before 
and after this correction as the peak phase shifts to 
earlier orbital phases as the super-orbital phase 
advances and the amplitude of the orbital light curve is 
maximized at $\varphi_{\rm s} = 0.5$.

\section{Basic model and numerical code for modeling light curves}

As discussed in Paper I, the light sources of the orbital 
signal consist of three components: (1) the orbital hump 
due to the bright spot produced by the collision of 
the gas stream with the disk at its rim, peaking 
at the orbital phase $\varphi_{\rm o}=0.8-0.9$, 
(2) the irradiation of the secondary by the tilted accretion 
disk, peaking around $\varphi_{\rm o}=0.5$, and 
(3) the ellipsoidal variation of the secondary star.  
As well known in CVs, the relative importance of the three 
components varies with the system brightness.  
When the system is faint, i.e., in the quiescence of 
a dwarf nova, the orbital hump usually dominates in 
the orbital light curve.  When the system brightens in 
the dwarf-nova outburst, the amplitude of the orbital hump 
in the orbital light curve decreases with the increase of 
the system brightness and the orbital hump almost 
disappears at the outburst maximum.  
This is because the orbital hump is swamped by the increased 
background radiation, i.e., the increased disk radiation.  
As discussed in Paper I and seen in Fig.~8 of Paper I, 
the same is true in the case of KIC 9406652 and 
the irradiation of the secondary dominates in the orbital 
light curve during the brightening stage.  
In our modeling, we do not consider the gas stream 
which produces the bright spot at the disk rim (i.e., 
the orbital hump) during the brightening stage.  

\begin{figure}[htb]
\vspace{-5mm}
\begin{center}
\FigureFile(80mm, 50mm){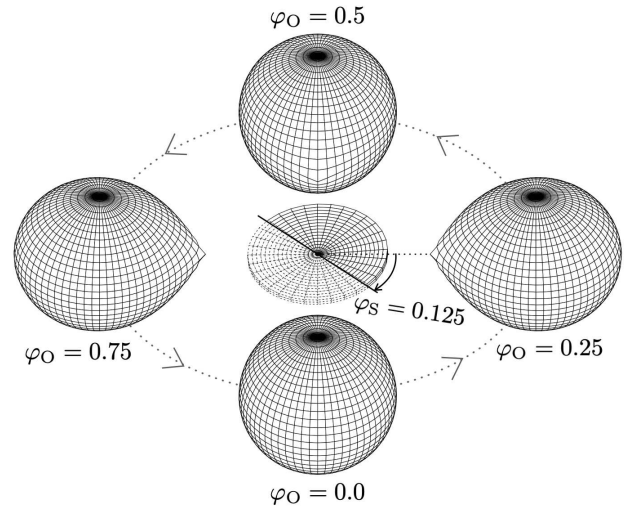}
\end{center}
\vspace{3mm}
\caption{
Schematic figure of our basic model geometry of KIC 9406652.  Here, the inclination angle is assumed to be 50 deg.  The central black point represents a WD and we put a tilted disk at the center with a tilt angle of 4.5 deg for visibility.  The nodal line is indicated by a diametric line passing the central WD.  The secondary star orbits counterclockwise around the central WD and the tilted disk precesses clockwise.   
We plot parts of the disk by the solid/dashed lines if its equatorial plane is above/below the orbital plane.  
The phases, $\varphi_{\rm s}$ and $\varphi_{\rm o}$, stand for that of the precession of the tilted disk (or that of the super-orbital modulation) and the orbital phase, respectively.  The phase zero for $\varphi_{\rm s}$ is defined when the nodal line of the tilted disk is perpendicular to the line of sight and the front face of the tilted disk turns towards the observers.  The super-orbital phase in this figure is 0.125.  The orbital phase zero is defined when the secondary star is at the inferior conjunction.  
}
\label{ellip-config}
\end{figure}

The basic model we use here is the standard semi-detached 
binary model of the cataclysmic variable star, consisting 
of the WD primary star, the Roche-lobe filling 
secondary star, and the accretion disk, which contribute 
to the orbital light curve during the brightening stage.  
The accretion disk is tilted out of the binary orbital plane 
with a fixed tilt angle, $\theta$, and its nodal line 
rotates retrogradely.  
We do not consider any warping of the disk plane and 
it is assumed to be coplanar.  The accretion disk is assumed 
to be circular and axisymmetric and it has a fixed outer 
radius.  
The secondary star is assumed to be irradiated by the bright 
tilted disk.  
The observer looks at this system with the inclination angle, 
$i$.  Figure \ref{ellip-config} illustrates our basic model.  

The numerical code that we use is the revised version of 
the code developed by \citet{hac01RN} who simulated light 
curves of recurrent novae.  
Although the method for calculating the flux is the same 
as that described in \citet{hac01RN}, we apply the response 
functions provided by \citet{man15filter} and 
the {\it Kepler} team\footnote{$<$https://nexsci.caltech.edu/workshop/2012/keplergo/CalibrationResponse.shtml$>$} in the revised code.  
In this numerical code, the surfaces of the accretion disk 
and the secondary star are divided into patches as displayed 
in Figure \ref{ellip-config}, and each patch emits photons 
by black body radiation at the local temperature.  
The numbers of surface patch elements are 16$\times$64 
(latitude $\times$ longitude) for each of the northern and 
the southern hemispheres of the secondary, 16$\times$32 
(radial $\times$ azimuthal) for the tilted disk as for 
each of the front face (seen from the observer) and 
the back face.  
We divide the surface patches for the sidewall of the disk 
by 2$\times$32 (upper and lower sides $\times$ azimuthal).  
The original code by \citet{hac01RN} can treat irradiation 
of these two components (i.e., the accretion disk and 
the secondary) by each other.  
In our study, we consider only the irradiation of the secondary 
by the accretion disk because the accretion disk is much brighter 
than the secondary star in the brightening stage of KIC 9406652.
The code also includes the emission from the WD, though 
it is negligible in our case.  
The numbers of surface patch elements for the WD is 
16$\times$32 (latitude $\times$ longitude).  
The details of this numerical code are described 
in \citet{hac01RN}.  

As for each surface element on the secondary star, 
the radiation from the accretion disk visible from 
the particular surface element is summed up and is used to 
heat the surface element.  
Here we introduce an efficiency parameter of the irradiation, 
$\eta$, in our code, in such a way that a fraction, $\eta$, 
of the irradiating energy is used to raise the local temperature 
of the particular surface element.  
That is, $\eta = 1.0$ if all energy received at each surface 
element by the irradiating radiation is used to raise 
the temperature of the surface element, and $\eta = 0$ if 
no heating occurs by the irradiation.  
In our modeling, $\eta$ is assumed to be a free adjustable 
parameter but it is assumed to take a given value 
once the inclination is specified.

\section{Parameters for modeling}

\subsection{Binary parameters}

We first reconsider the binary parameters of KIC 9406652 
according to \citet{gie13j1922} and \citet{kim20kic940}.  
We set the orbital period ($P_{\rm orb}$) to be 
0.2545094~d, a value derived in \citet{kim20kic940}.  
\citet{gie13j1922} provided the radial velocity curves of 
the WD and the secondary star in this object.  
We have fitted them with sinusoidal curves with the fixed 
orbital period and have obtained the mass ratio ($q \equiv 
M_2/M_1$) to be 0.87$\pm$0.08.  
Here $M_1$ and $M_2$ are the masses of the primary and 
the secondary stars.  
We thus set the mass ratio to be 0.87 in this study.
If the orbital period is well determined, the mass of 
the Roche-lobe filling secondary star is constrained 
by the empirical relation investigated by 
\citet{kni06CVsecondary} as discussed by \citet{gie13j1922}.  
We set the secondary mass to be 0.73~$M_{\solar}$ 
from this consideration and 
therefore the WD mass to be 0.84~$M_{\solar}$.  
The inclination angle is constrained by the following mass 
function, that is, 
\begin{equation}
\frac{P_{\rm orb}}{2\pi G} = \frac{M_2 \sin{i}^3}{q(1+q)^2 K_2^3}, 
\label{mass-function}
\end{equation}
where $G$ is the gravitational constant and 
$K_2$ is the semi-amplitude of the radial velocity curve 
of the secondary star.  The orbital inclination 
is estimated to be $\sim$50~deg from 
equation (\ref{mass-function}). 
Its possible range determined from 
the spectroscopic observations is then 45--60 deg, 
if we take into account possible uncertainties in 
$M_2$, $K$, and $q$.  
The binary separation ($a$) is calculated from 
$4 \pi^2 a^3 = P_{\rm orb}^2 G (M_1 + M_2)$ and 
it is estimated to be 1.37$\times$10$^{11}$~cm.  

We next set the effective temperature of the WD and 
the secondary star.  
The mass accretion rate onto the WD in KIC 9406652 
is thought to be high in the brightening stage.  
The WD temperature ($T_1$) would become as high as 
$\sim$40,000~K in the dwarf-nova outburst (see Table 2.8 
in \citet{war95book} and \citet{god17ugem}).  
We, therefore, set $T_1$ to be 40,000~K, though its exact 
value is not important since the emission from the WD is 
negligible in the optical wavelength range of our interest 
as described in section 3.
As for the effective temperature of the secondary star 
($T_2$), we \textcolor{black}{first} considered a possibility to determine it 
from the infrared measurements referred in \citet{gie13j1922}.  
However, it is difficult to do this 
because the disk is bright even at infrared wavelengths.  
On the other hand, according to the empirical relation between 
the spectral type of the secondary star and the orbital 
period investigated in \citet{kni06CVsecondary}, $T_2$ would 
be estimated to be around 4,370~K from the orbital period of 
KIC 9406652 and its dispersion is a few hundred Kelvin.  
That is, $T_2 \sim$~4,100--4,600~K is allowed.  
As will be discussed in the next subsection, the temperature 
of the secondary star, $T_2$ should rather be determined 
from the flux ratio of the secondary star to the disk 
and its determination is postponed to the next subsection.

\subsection{Parameters of the accretion disk}

We consider a tilted disk with a tilt angle $\theta$.  
Its temperature distribution is assumed to be that of 
the standard disk expressed as 
\begin{eqnarray}
T = T_{*} \left(\frac{r}{r_{\rm in}} \right)^{-3/4}\left[1-\sqrt{\frac{r_{\rm in}}{r}} \right]^{1/4},~
T_{*} = \left(\frac{3 G M_1 \dot{M}}{8 \pi \sigma r_{\rm in}^{3}} \right)^{1/4}. 
\label{temp-steady}
\end{eqnarray}
Here, $r$, $r_{\rm in}$, $G$, $M_1$, $\dot{M}$ and $\sigma$ 
represent the distance from the central object to a given ring 
of the disk, the innermost radius of the disk, 
the gravitational constant, 
the WD mass, the mass accretion rate at a given radius, and 
the Stefan-Boltzmann constant, respectively \citep{sha73BHbinary}.  
We also set the innermost radius of the disk and the WD radius 
to be 6.0$\times$10$^{8}$~cm because the WD mass is about 
0.84$M_{\solar}$.  
We assume that the outermost radius of the accretion disk is 
given by the tidal truncation radius because we study 
the orbital light curves during the brightening stage. 
The tidal truncation radius is calculated to be 0.328 
in units of the binary separation from the results of 
\citet{pac77diskmodel} for the binary system with 
the mass ratio $q = 0.87$.  

Let us now discuss the mass transfer rate and the temperature 
of the secondary star in our modeling. 
IW And stars are supposed to be the very cataclysmic 
binary systems near the borderline between dwarf novae 
and nova-like variables and thus their mass transfer rate 
is supposed to be near the critical mass transfer rate 
($\dot{M}_{\rm crit}$) above which the disk is thermally stable.  
In the case of the steady standard disk, the mass-transfer rate 
is equal to $\dot{M}$ in equation (\ref{temp-steady}).  
Therefore $\dot{M}$ is estimated to be 10$^{17.66}$~g~s$^{-1}$ 
for the binary parameter of KIC 9406652 (see, \cite{ham98diskmodel}).  

In the case of KIC 9406652, the distance is accurately 
measured to be 348~pc by the {\it Gaia} parallax 
\citep{bai18gaia}.  
The $V$-band photometry is performed by the All-Sky 
Automated Survey for Supernovae (ASAS-SN) data 
\citep{dav15ASASSNCVAAS}.  
On the other hand, the flux ratio, $F_2/F_1$, 
for that of the secondary to that of the disk, was derived 
to be 0.06(2) through the {\it Kepler} response function 
in Paper I based on the result in \citet{gie13j1922}.  
It is noted here that this value is derived from 
optical spectra taken in the quasi-standstill state.  
We now explore the best set of $\dot{M}$ and $T_2$ to satisfy 
the following two conditions: (1) the $V$-band magnitude is 
around 11.5--11.9 on average and (2) $F_2/F_1$ = 0.06(2) 
in quasi-standstills.  
We have run the numerical code for this purpose and 
have found that $\dot{M}$ = 10$^{17.7}$~g~s$^{-1}$ and 
$T_2$ = 4,100~K satisfy these conditions over $\eta$ = 
0.0--1.0 and $i$ = 30--55~deg for the distance of 
KIC 9406652.  
The mass transfer rate thus obtained is found to be very 
near the critical mass transfer rate mentioned above.  
Since the $V$-band flux is brighter by $\sim$0.3 mag in 
the brightening phase than that in quasi-standstills,  
$\dot{M}$ will be 10$^{17.9}$--10$^{18.0}$~g~s$^{-1}$ 
in the brightening stage.  
We therefore set $\dot{M}$ to be 10$^{18.0}$~g~s$^{-1}$ 
and $T_2$ = 4,100~K in our modeling during 
the brightening stage.  

We postulate that the thickness of the disk basically 
follows that of the standard disk.  
The equation that we used is represented as 
\begin{eqnarray}
\frac{H}{r} = 1.72 \times 10^{-2} \alpha^{-1/10} \dot{M}_{16}^{3/20} \left({\frac{M_1}{M_{\odot}}}\right)^{-3/8} r_{10}^{1/8} \left[1-\sqrt{\frac{r_{\rm in}}{r}} \right]^{3/5},  
\label{height-steady}
\end{eqnarray}
where $H$, $\dot{M}_{16}$ and $r_{10}$ represent the height of 
the disk, $\dot{M}/10^{16}$~g~s$^{-1}$ and $r/10^{10}$~cm, 
respectively \citep{sha73BHbinary}.  
The $\alpha$ value is assumed to be 0.2.  
The thickness of the disk with our binary parameters is then 
given by $H/r \sim 0.05$ at the outer edge.  
On the other hand, we may need to integrate the vertical disk 
structure instead of the one-zone approximation to accurately 
know the thickness of the disk and \citet{sma92CVdisk} showed 
the disk in the outburst state has $H/r \sim 0.1$ at the disk 
outer edge.  
Since the standard disk is thinner than this, we also study 
the case of a thicker disk with $H/r = 0.1$ at the disk 
outer edge by multiplying a constant factor to equation 
(\ref{height-steady}).

\section{Modeling of the orbital light curve during the brightening stage}

\subsection{Our approach to the problem}

We try to simulate the orbital light curves during 
the brightening stage of KIC 9406652, i.e., 
Figure \ref{orb-b} and the right figure of Figure \ref{orb8b} 
on the basis of our model of the tilted disk and 
the irradiated secondary star.  
The purpose is to examine how well our model can reproduce 
the variation of the observed orbital light curves with 
respect to the super-orbital phase.  
The parameters used in our modeling are, 
(1) the irradiation efficiency, $\eta$, 
(2) the orbital inclination, $i$, and 
(3) the tilt angle, $\theta$.  
As discussed in subsection 4.1, the orbital inclination of 
KIC 9406652 is constrained to be 45--60 deg from 
the spectroscopic considerations. 
However, we study light curves for a much wider range 
in the orbital inclination with 30--60 deg and try to see 
if any preference in the orbital inclination exists from 
the photometric considerations.  
As for the tilt angle, $\theta$, we have concluded 
in Paper I that it was lower than 3 deg from 
the semi-amplitude of super-orbital light modulations 
in KIC 9406652.  
We here perform numerical modeling of the observed orbital 
light curves for a much wider range in the tilt angle with 
0.5--4.5 deg as well. 
As will be shown in subsection 5.2, the best value of 
the irradiation efficiency, $\eta$, is determined only 
by the inclination and it is almost independent 
of the tilt angle.  
The free parameters to be explored are thus the orbital 
inclination and the tilt angle.  

In what follows, we perform numerical modeling 
for various orbital inclination and tilt angles.  
The two different effects produce variations in the orbital 
light curves, that is, variations in the projected area of 
the tilted disk and those in the irradiation of the secondary star.  
The former affects the mean light level of the orbital light 
curve with different super-orbital phases and the latter affects 
variations of the orbital light curve profile with different 
super-orbital phases and these effects are separable.  
The former effect can be measured by the semi-amplitude of 
the super-orbital light modulations.  
The latter effect can be measured by the comparison of 
the observed light curve profiles with the simulated ones. 
In order to judge which parameters can reproduce 
the observed light curves the best, we use the following 
two criteria.  
The first criterion is the semi-amplitude of super-orbital 
light modulations, which will be discussed in subsection 5.3. 
The second one is the goodness of fit between the observed 
light curves and the model light curves, represented 
by $\chi^2$, the details of which will be explained 
in subsection 5.6.  

We \textcolor{black}{first} discuss the efficiency parameter, $\eta$, in 
subsection 5.2, and we then treat the observational light 
curve of super-orbital light modulations in subsection 5.3.  
We present a typical example of our modeling for 
the orbital light curves with eight super-orbital phases, 
$\varphi_{\rm s}$, in subsection 5.4 and we discuss 
the dependence of our results on the tilt angle, 
$\theta$, and the orbital inclination, $i$, in subsection 5.5.  
We finally compare simulated light curves for various $i$ and 
$\theta$ by using the two criteria mentioned above to see 
which parameter set can reproduce the observed light curves 
the best.

\subsection{Determination of the irradiation efficiency, $\eta$}

\begin{figure}[htb]
\begin{center}
\FigureFile(80mm, 50mm){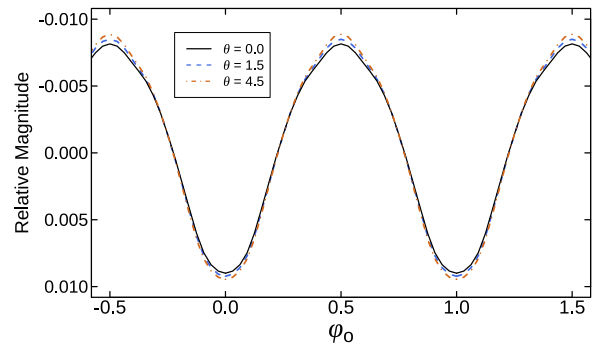}
\end{center}
\caption{Averaged orbital signals of the tilted disk of $\theta$ = 1.5 and 4.5 deg with that of the non-tilted disk.  We depict the dashed and dot-dashed lines by averaging the calculated orbital signals of the tilted disk with 8 orientations corresponding to $\varphi_{\rm s}$ = 0.0, 0.125, 0.25, 0.375, 0.5, 0.625, 0.75, and 0.875.  The vertical axis is the relative magnitude through the {\it Kepler} response function.  
}
\label{check-eta}
\end{figure}

The determination of the irradiation efficiency factor 
is important since the irradiation of the secondary 
star by the tilted disk plays a major role in our model.  
In our study, we treat the efficiency factor, $\eta$, 
to be an adjustable free parameter. 
We use the orbital light curve of Figure \ref{orb-b} 
which is obtained by using all data during the brightening 
stage for this purpose.  
It has turned out from our modeling that an orbital light 
curve generated by the superposition of light curves at 
various orientations of a tilted disk does not depend 
very much on the tilt angle, $\theta$, and it is basically 
given by the light curve of a non-tilted disk.  
Figure \ref{check-eta} displays the averaged 
orbital signals of the tilted disk of $\theta$ = 1.5 and 
4.5 deg together with that of the non-tilted disk.  
There are only tiny differences between them.
Since the orbital light curve of Figure \ref{orb-b} is 
obtained by all data during the brightening stage, 
it represents the phase averaged orbital light curve of 
the tilted disk plus the irradiated secondary star and 
we can compare Figure \ref{orb-b} with the simulated light 
curve of the non-tilted disk for a given orbital 
inclination, $i$.  
We can estimate the best fit value of $\eta$, for various 
inclinations ranging from 30 to 60 deg with a step of 5 deg and 
the result is summarized in Table \ref{i-eta2}.  
Here we determine the best-fitting parameter, $\eta$, 
for a given inclination by calculating $\chi^2$ between 
the observed light curve of Figure \ref{orb-b} and 
model light curves with various $\eta$, with accuracy 
in 3 figures.  
In what follows, we use the $\eta$ values given in 
Table \ref{i-eta2} in our modeling, once the orbital 
inclination is specified.  

\begin{table*}
	\caption{Relation between the inclination angle ($i$) in units of degree and the efficiency of irradiation ($\eta$).  }
	\vspace{2mm}
	\label{i-eta2}
	\centering
	\begin{tabular}{|c|ccccccc|}
\hline
$i$ & 30 & 35 & 40 & 45 & 50 & 55 & 60\\
\hline
$\eta$ & 0.577 & 0.477 & 0.399 & 0.337 & 0.286 & 0.241 & 0.191\\
\hline
\end{tabular}
\end{table*}

\begin{figure*}[htb]
\begin{center}
\begin{minipage}{0.49\hsize}
\FigureFile(80mm, 50mm){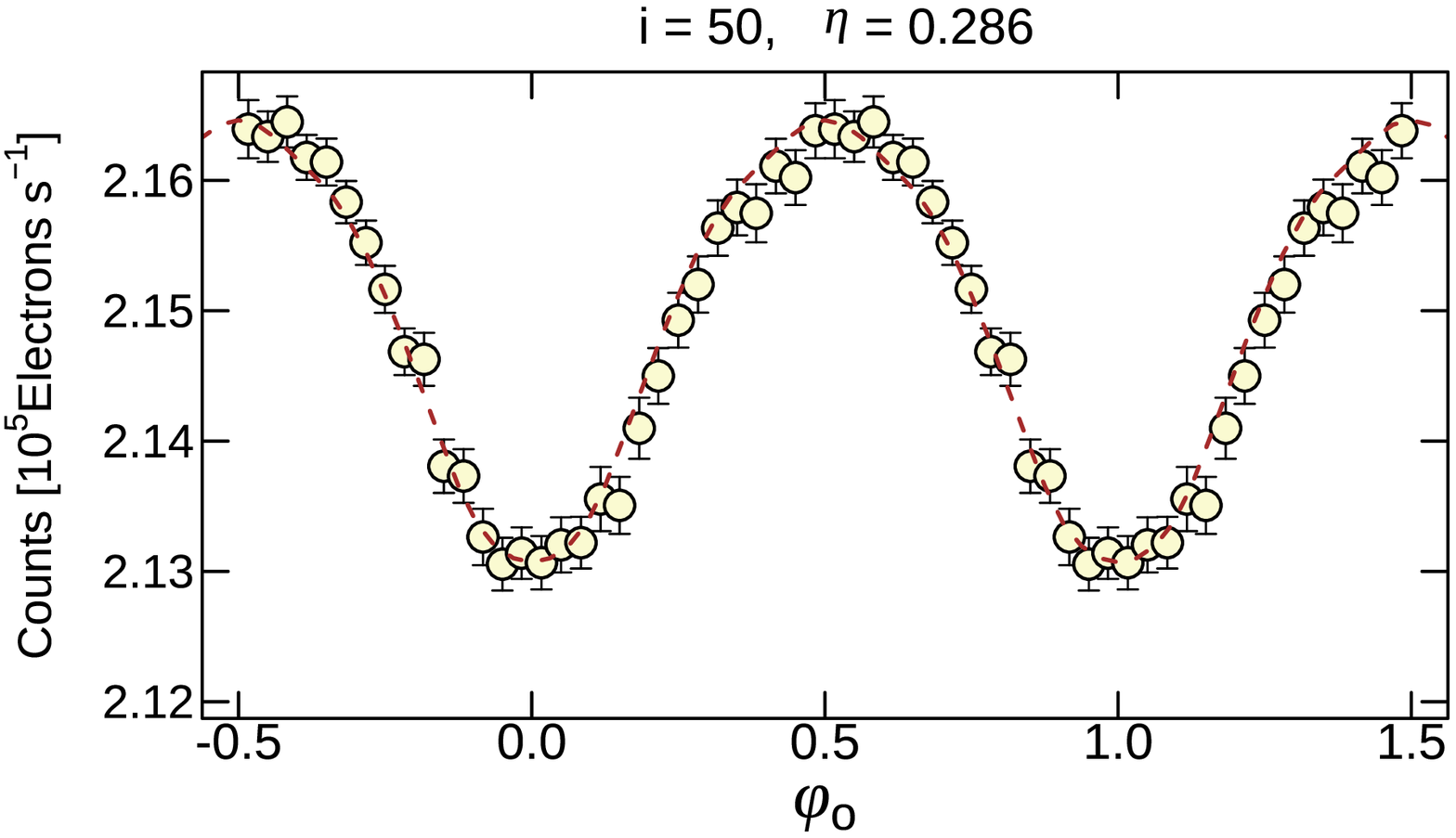}
\end{minipage}
\begin{minipage}{0.49\hsize}
\FigureFile(80mm, 50mm){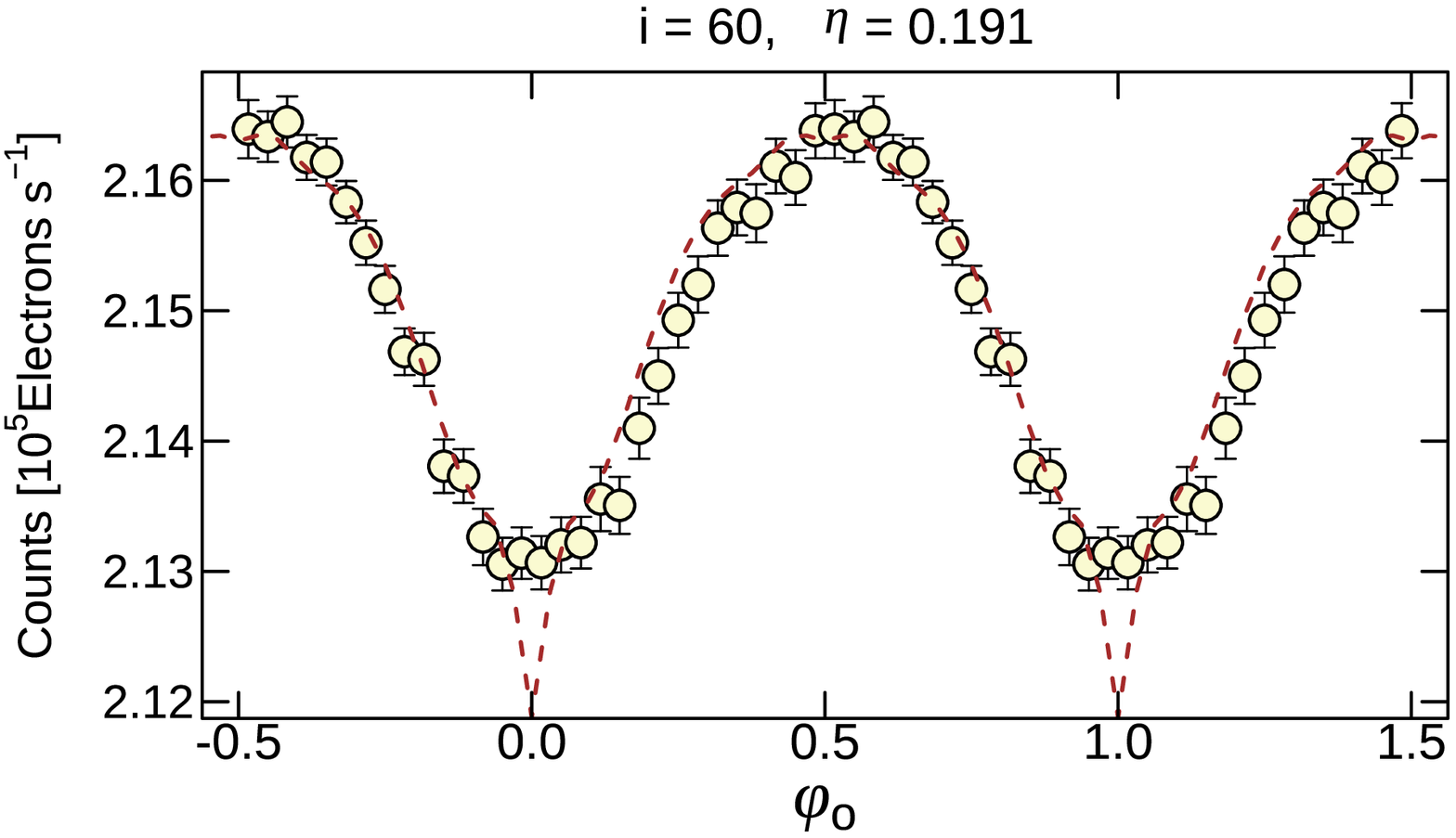}
\end{minipage}
\end{center}
\caption{Model light curves compared with the observational light curve in Figure \ref{orb-b}.  The dashed line shows the model light curve while the circles with error bar are those of observations.  }
\label{fig8model}
\end{figure*}

The best-fitting model light curves for $i$ = 50 and 60 deg 
are displayed in Figure \ref{fig8model} together with 
the observational light curve of Figure \ref{orb-b}.  
As seen in the left panel in Figure \ref{fig8model}, 
the model light curve fits very well with the observational data.
On the other hand, the model light curve for $i$ = 60~deg 
in its right panel deviates from the observational light curve 
in such a way that a sharp dip due to the eclipse of 
the outer disk by the secondary is clearly visible around 
$\varphi_{\rm o}$ = 0.0 in the model light curve, 
while the observed light curve does not show any dip but 
a rounded minimum. 
We, therefore, exclude the case of $i$ = 60 deg safely, and 
hereafter we do not consider this case anymore.  
\textcolor{black}{
We confirm that the best-fitting model light curves for all 
$i$ except for 60 deg do not show any big differences 
between the model light curve of $i$ = 50 deg shown 
in the left panel of Figure \ref{fig8model} and the others.  
}

\subsection{Light curve of super-orbital modulations and its semi-amplitude}

\begin{figure}[htb]
\begin{center}
\FigureFile(80mm, 50mm){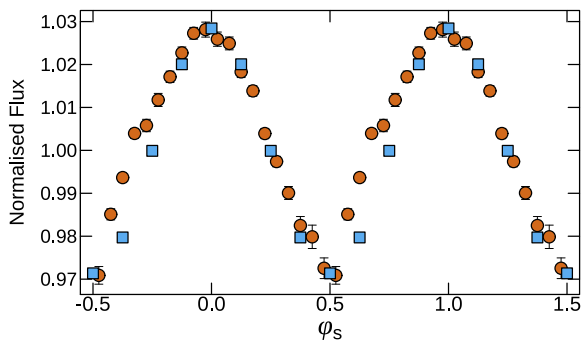}
\end{center}
\caption{Phase-averaged profile of super-orbital signals during brightening in KIC 9406652 (the circles) and the prediction from our model light curves of $i$ = 50~deg and $\theta$ = 1.5~deg (the squares).  The latter is derived by normalized and averaging each model light curve displayed in the left panel of Figure \ref{basic-i50-t1.5}.  }
\label{sup-b}
\end{figure}

We construct the light curve of super-orbital modulations 
during the brightening stage.  
To do so, two short-period signals (i.e., the orbital and 
the negative superhump signals) are removed from 
the original light curve in advance as done in subsection 3.4 
of Paper I.  
We then divide those data into intervals with the length of 
4.13~d, i.e., the average period of super-orbital signals, 
perform locally-weighted polynomial regression (LOWESS: 
\cite{LOWESS}) to evaluate the intrinsic variation 
originating from the accretion process for each interval, 
and subtract it from the light curve.  
We next fold the light curve by the 4.13-d period 
and fit the folded one by a sinusoidal curve and 
set the epoch as the light maximum is $\varphi_{\rm s}$ 
= 0.0.  
The resulting light curve is shown in Figure \ref{sup-b}.  
We can construct the corresponding model light curve 
from our modeling.  As shown in the next subsection, 
we construct eight orbital light curves with eight 
super-orbital phases, $\varphi_{\rm s}$ = 0.0, 0.125, 0.25, 
0.375, 0.5, 0.625, 0.75, and 0.875 for the case of 
$i$ = 50~deg and $\theta$ = 1.5 deg corresponding to 
the right figure of Figure \ref{basic-i50-t1.5}.  
By averaging each light curve, we obtain 8 corresponding 
fluxes and we show them by blue squares in Figure \ref{sup-b}.  

If the super-orbital modulation is produced by the variation 
in the projected area of the tilted disk with the nodal 
precession, the expected light curve must be sinusoidal 
(see, the blue squares in Figure \ref{sup-b}).  
However, the observed light curve shown in Figure \ref{sup-b} 
deviates from the sinusoidal form.  
This indicates that the intrinsic variation of the disk 
luminosity is not properly subtracted from the light curve 
and some contamination remains.  
Since the timescale of intrinsic light variations 
particularly in the brightening stage is of the order of 
a few days, the contamination to super-orbital modulations 
from the intrinsic light variation of the accretion disk is 
inevitable and no good way exists to subtract this effect 
completely.  
\textcolor{black}{
The intrinsic light variation could lead to either 
an overestimate or an underestimate in the semi-amplitude 
of super-orbital modulations.}  
We must take into account this problem in mind.  
The semi-amplitude of normalized super-orbital signals 
during brightening is $\sim$0.03 from the orange circles 
of Figure \ref{sup-b}, which corresponds to the tilt angle 
to be $\sim$1.6 deg for the case of $i$ = 50 deg 
(see equation (5) in Paper I).  

\textcolor{black}{
We here discuss another possibility leading to an underestimate 
in the semi-amplitude of super-orbital modulations.}  
The amplitude of the orbital signal for $\varphi_{\rm s} 
\sim 0.5$ is bigger than that for $\varphi_{\rm s} \sim 0.0$ 
(see also Figure \ref{orb8b}) because the observer looks 
at the wider high-temperature region of the irradiated surface 
of the secondary star at $\varphi_{\rm s} \sim 0.5$, i.e., 
the irradiation effect seems to be stronger to the observer.  
The increase \textcolor{black}{in} the mean flux level by 
the irradiation effect is larger at $\varphi_{\rm s} \sim 0.5$ 
than that at $\varphi_{\rm s} \sim 0.0$ 
because higher-amplitude variations are superimposed at 
$\varphi_{\rm s} \sim 0.5$.  
\textcolor{black}{
This causes the semi-amplitude of super-orbital signals 
to be underestimated.}  
The difference in the normalized semi-amplitude is about 0.0036 
between the orbital signal of the top panel of the left column 
in the right figure of Figure \ref{orb8b} and that of 
the top panel of the right column in the same figure.  
The semi-amplitude of super-orbital signals is possibly 
underestimated by \textcolor{black}{the amount of} this difference.

\subsection{Typical example of our modeling for orbital light curves with various super-orbital phases}

\begin{figure*}[htb]
\begin{center}
\setlength{\fboxrule}{0.7pt}
\begin{minipage}{0.49\hsize}
\fbox{
\FigureFile(80mm, 50mm){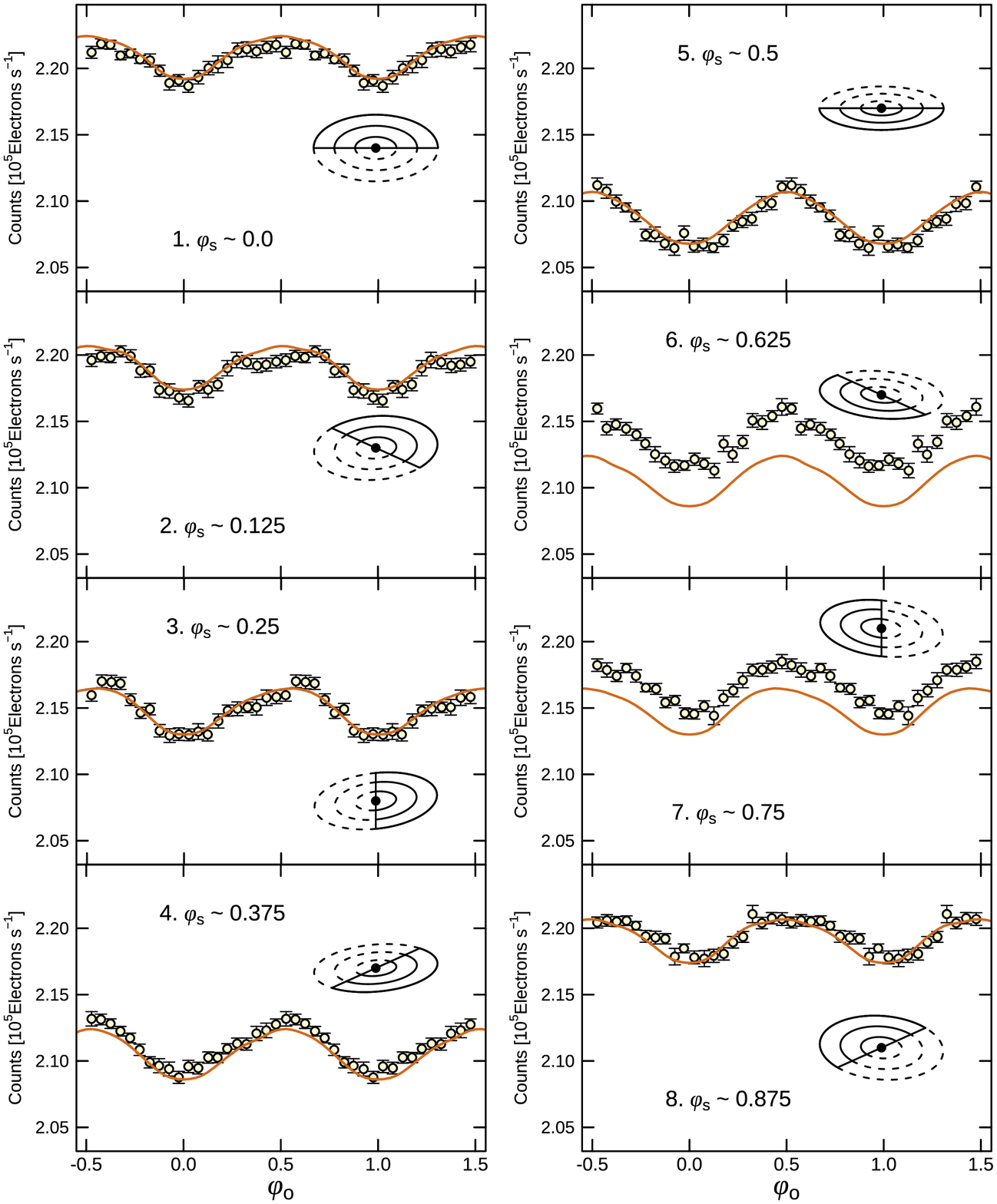}
}
\end{minipage}
\begin{minipage}{0.49\hsize}
\fbox{
\FigureFile(80mm, 50mm){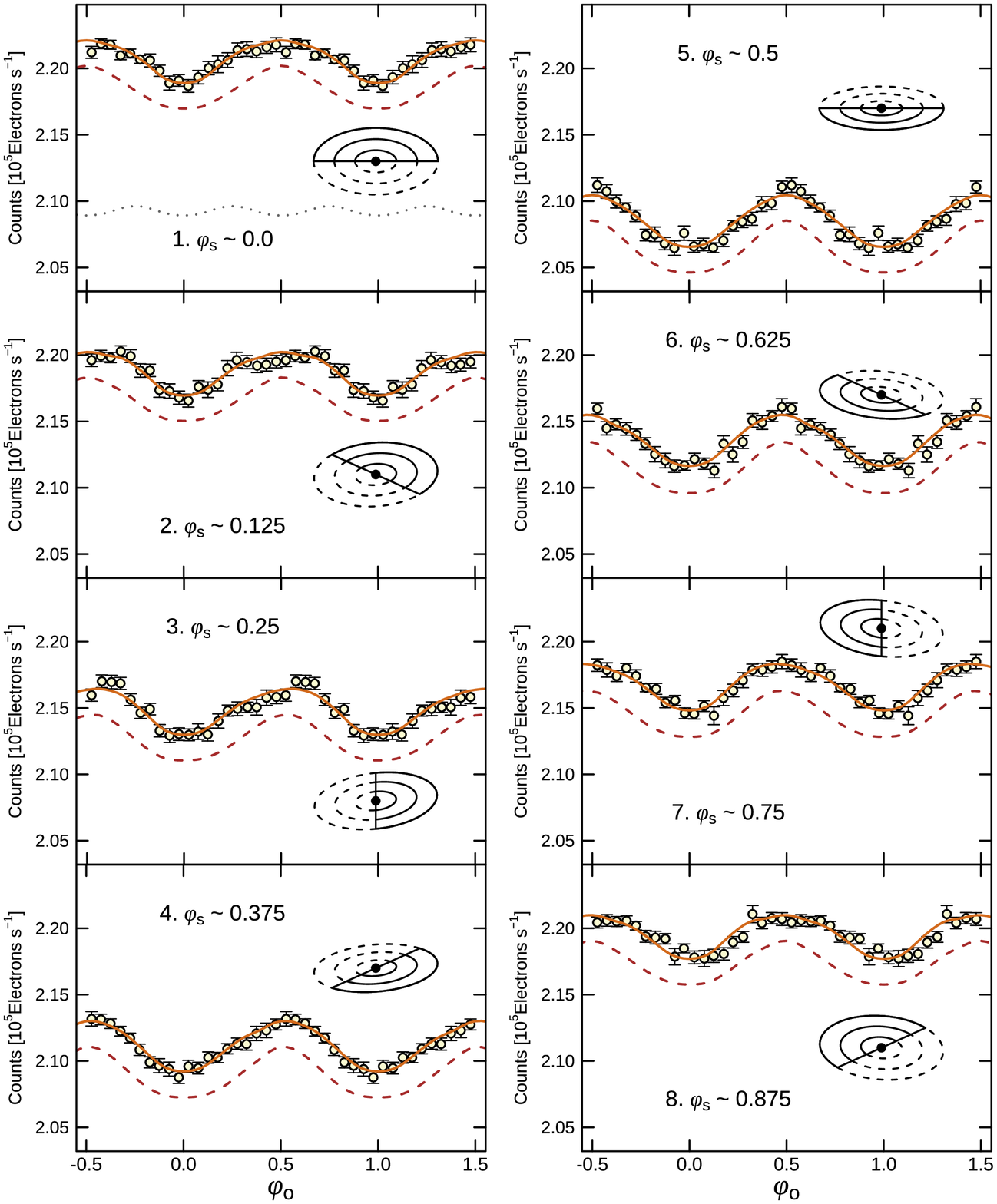}
}
\end{minipage}
\end{center}
\caption{Comparison of model light curves (red solid lines) with observed ones (dots with error bars) shown in the right figure of Figure \ref{orb8b} for the case of $i$ = 50~deg and $\theta$ = 1.5~deg.  The eight left-hand side panels illustrate the case when the mean flux level of all model light curves is adjusted to that of the averaged orbital light curve of all super-orbital phases, while the eight right-hand side panels do the case in which the flux level of each model light curve at a given $\varphi_{\rm s}$ is adjusted to that of the corresponding observed light curve for the purpose of a direct comparison of light curve profile.  The model light curve in each panel of the right-hand side is decomposed into two different effects: the irradiation effect and the ellipsoidal modulation of the secondary, and they are shown by dashed and dot lines.  The ellipsoidal light curve is displayed only in the top left panel of the right-hand side figures below the irradiation light curve (red dashed curve) because it is the same for all eight $\varphi_{\rm s}$.  The irradiation and ellipsoidal light curves are shifted by 0.06$\times$10$^5$ and 2.01$\times$10$^5$ for visibility.  }
\label{basic-i50-t1.5}
\end{figure*}

Here we describe our simulation method to construct 
an orbital light curve at a fixed super-orbital phase.  
We \textcolor{black}{first} specify the orbital inclination and 
the tilt angle of the disk (in this example, 
$i$ = 50 deg and $\theta$ = 1.5 deg).  
We then specify a particular super-orbital phase 
(for instance, for the case of $\varphi_{\rm s}$ = 0.125 
in Figure \ref{ellip-config}) where the zero point of 
the super-orbital phase is defined when the tilted disk 
is directly facing the observer.  
We fix the orientation of the tilted disk in this position 
but the secondary star orbits around the WD and the tilted 
disk (counterclockwise as shown in Figure \ref{ellip-config}).  
We calculate all \textcolor{black}{fluxes} from the tilted accretion disk, 
the central WD, and the irradiated secondary star 
at a given orbital phase by the method described 
in section 3. 
We now construct the orbital light curve for a specified 
super-orbital phase by calculating \textcolor{black}{fluxes} for various 
orbital phases.  
We then advance the super-orbital phase by 0.125 by changing 
the orientation of the tilted disk due to the nodal precession 
in the retrograde direction (for example, from 0.125 to 0.25 
clockwise in Figure \ref{ellip-config}). 
We finally obtain eight orbital light curves corresponding 
to eight super-orbital phases and we illustrate our result 
in the left-hand 8 panels of Figure \ref{basic-i50-t1.5} 
for the case of $i$ = 50~deg and $\theta$ = 1.5~deg.  

We find from the eight panels on the left of 
Figure \ref{basic-i50-t1.5} that the simulated light 
curves fit fairly well with those observed ones except that 
the mean light level of the simulation does not agree well 
with that of observation for some light curves 
(particularly for the cases of $\varphi_{\rm s}$ 
= 0.625 and 0.75 in the sixth and seventh panels of 
the left-hand side).  
As discussed in subsection 5.3, this is mainly caused 
by our deficiency in subtracting properly the influence of 
the intrinsic variation from the observed light curves 
and it is not any deficiency of our modeling.  
We here compare only the observed and calculated values 
for the semi-amplitude of the super-orbital light signals 
as our criterion 1.  

As for the second criterion, we rather plot each orbital 
light curve at a given super-orbital phase by adjusting 
the mean flux level of the simulated light curve to that of 
the observational one.  
The eight orbital light curves in the right-hand side of 
Figure \ref{basic-i50-t1.5} exhibit our results 
thus obtained.  
We see that our simulated light curve profiles reproduce 
the observed ones fairly well.  
That is, the most conspicuous observational features are 
reproduced by our model as the peak of the light curve shifts 
to earlier orbital phase with the advance of the super-orbital 
phase and the amplitude of orbital light curve is maximized 
at the super-orbital phase $\varphi_{\rm s} = 0.5$, 
and the light minima always occur at $\varphi_{\rm o} = 0.0$.  

The simulated orbital light curve profiles are produced by 
two different effects, that is, the ellipsoidal variation of 
the secondary star and variations of the irradiated secondary 
by the tilted disk.  
These two effects are shown separately in the right-hand 
eight panels of Figure \ref{basic-i50-t1.5}.  
Since the ellipsoidal effect of the secondary star is 
independent of the orientation of the tilted disk, 
we show it only in the first panel of the right-hand side 
figure corresponding to $\varphi_{\rm s}$ = 0.0.  
As well known, two peaks appear at $\varphi_{\rm o}$ = 0.25 
and 0.75 by the ellipsoidal effect.  
The amplitude of the simulated orbital light curve 
at $\varphi_{\rm s}$ = 0.5 seems to be smaller than 
the observed one.  
It is however difficult to judge \textcolor{black}{by eye} how well 
the simulated light curve profiles reproduce the observed ones 
and we, therefore, use criterion 2 in subsection 5.6.

\subsection{Dependence of our modeling on the tilt angle, $\theta$, and the orbital inclination, $i$}

\begin{figure}[htb]
\begin{center}
\FigureFile(80mm, 50mm){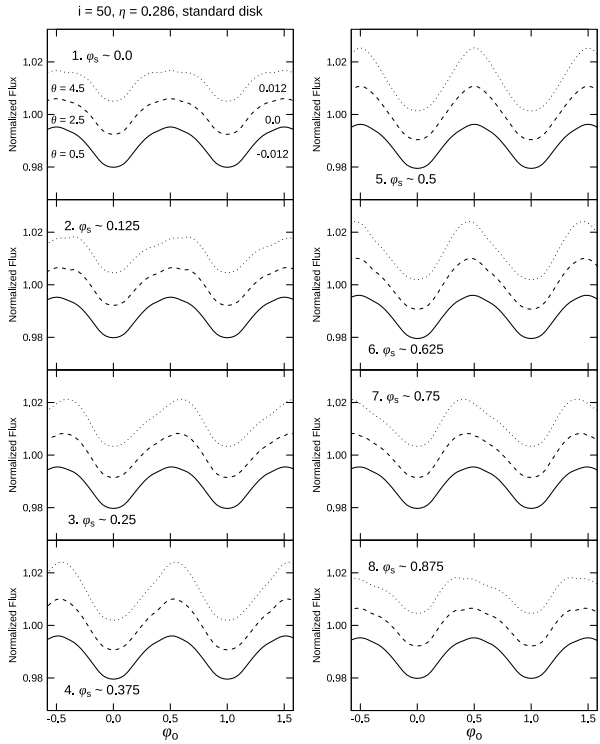}
\end{center}
\caption{Dependence of model light curves on the tilt angle for $i$ = 50~deg.  The tested tilt angles ($\theta$) are \textcolor{black}{0.5, 2.5, and 4.5~deg} and each resulting orbital profile is shifted in the vertical axis for visibility (see the left top panel).  }
\label{model-i50}
\end{figure}

Before evaluating all of the results of our modeling by 
the two criteria that we define, 
we first examine the dependence of the results 
on the tilt angle, $\theta$, by fixing the orbital 
inclination.  
Figure \ref{model-i50} illustrates our results for 
the case with $i$ = 50 deg.  We see that the amplitude of 
the orbital light curve for $\varphi_{\rm s}$ = 0.5 
clearly increases with an increase in the tilt angle 
while it does not for $\varphi_{\rm s}$ = 0.0.  
In other words, their ratio increases with the increase 
in $\theta$.  
We also find that the higher the tilt angle is, 
the larger the shift of the peak of the orbital light curve 
is with the advance in the super-orbital phase.  
We also see that the ellipsoidal effect becomes 
more important relative to the irradiation effect 
for $\varphi_{\rm s}$ = 0.0, as the tilt angle becomes higher, 
while the opposite is the case for $\varphi_{\rm s}$ = 0.5.  

\textcolor{black}{
We have also examined the dependence of our results 
on the orbital inclination for a fixed tilt angle.  
We find that the ellipsoidal effect of the secondary 
is most clearly visible for the highest orbital 
inclination as the two shoulders around the light maximum 
(for the cases at $\varphi_{\rm s}$ = 0.0 and 0.5) and 
that it becomes less clear with a decrease in 
the orbital inclination.  
Besides, the shift to an earlier orbital phase of 
the light maximum with the advance in $\varphi_{\rm s}$ 
becomes clearer, the lower the inclination is.  
}

\begin{figure*}[htb]
\begin{center}
\FigureFile(160mm, 50mm){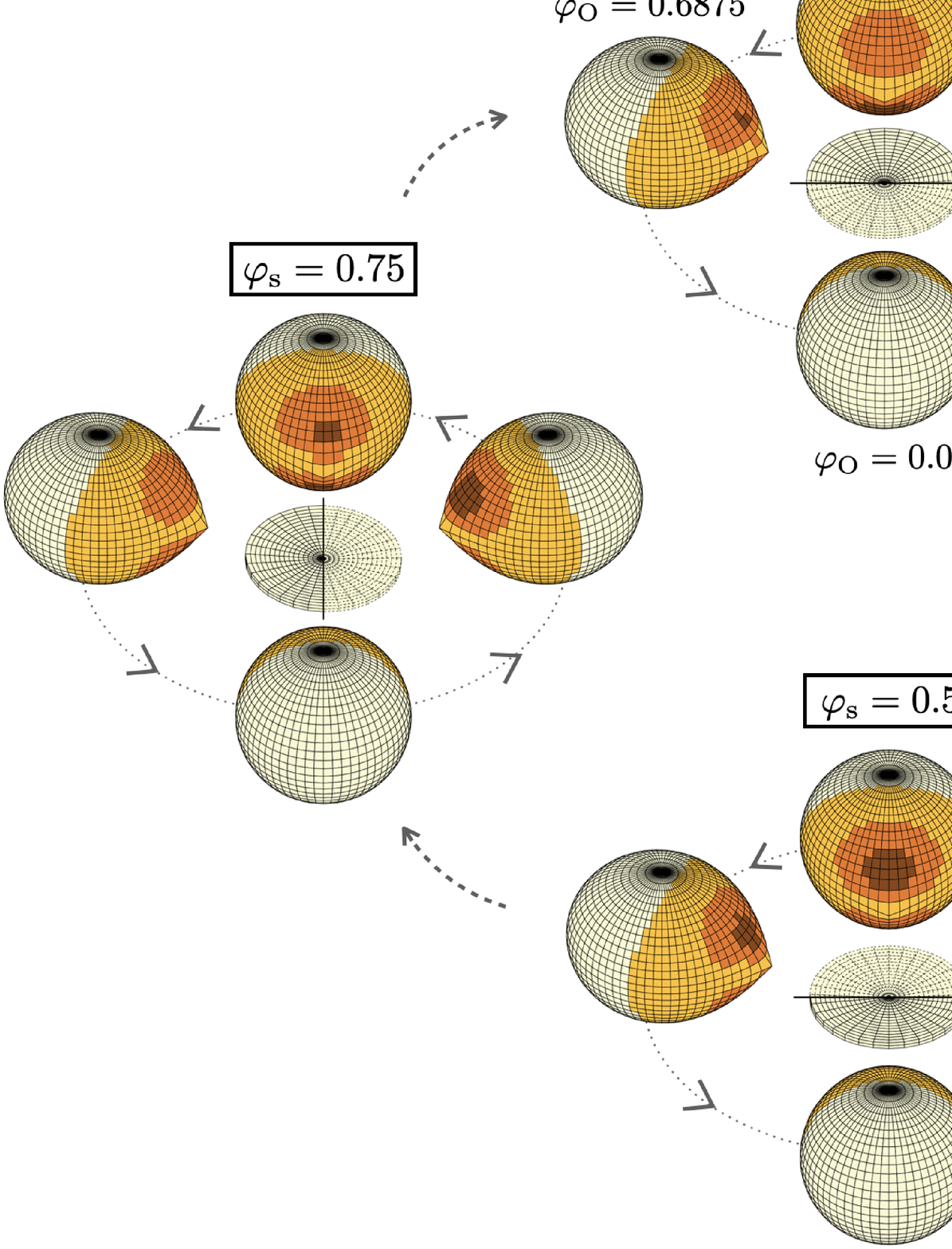}
\end{center}
\vspace{3mm}
\caption{Binary configurations with four super-orbital phases showing the tilted disk and the irradiated secondary star by colors in the case of $i$ = 50 deg and $\theta$ = 2.0 deg.  The temperature of the irradiated surface of the secondary star is shown in color, i.e., brown for that above 5,000~K, orange for that between 4,500~K and 5,000~K, and yellow for that below 4,500~K but above 4,100~K.  The super-orbital phase advances in the clockwise direction starting from the top figure for $\varphi_{\rm s}$ = 0.0 (where the tilted disk is facing toward the observer) and then 0.25, 0.5, and finally 0.75.  The secondary star makes one orbital revolution around the WD and the tilted disk counterclockwise in each frame, showing the four orbital phases at $\varphi_{\rm o}$ = 0.0, 0.3125, 0.5, and 0.6875.  The nodal line of the tilted disk is shown by the diametric line and the solid/dashed parts of the disk represent those for the above/below the orbital plane.  }
\label{irradiation-color}
\end{figure*}

We illustrate the binary configuration of the irradiated 
secondary star and the tilted disk for four super-orbital 
phases in Figure \ref{irradiation-color} 
to help for understanding our modeling.  
By comparing the illustration at $\varphi_{\rm s}$ = 0.0 
(the top panel) and that at $\varphi_{\rm s}$ = 0.5 
(the bottom panel), we see that the irradiated hemisphere 
of the secondary star is more easily observed at 
$\varphi_{\rm s}$ = 0.5 when the secondary star is 
at the superior conjunction and therefore the amplitude of 
the orbital light curve is maximized at this super-orbital 
phase of $\varphi_{\rm s}$ = 0.5.    
The temperature distribution of the irradiated surface 
is asymmetric with respect to the orbital phase 0.5 at 
$\varphi_{\rm s}$ = 0.25 and 0.75 (see the right and left 
panels), so that the light maximum of the orbital light curve 
deviates from the orbital phase 0.5.  
The orbital phase at the light maximum thus shifts towards 
the earlier orbital phase with the advance of 
the super-orbital phase.  

As for the dependence of light curves on the tilt angle 
of the disk, it is quite natural to understand that 
the above-mentioned effects are more enhanced 
if the tilt angle becomes higher.  
On the other hand, the same is true if the inclination 
angle becomes lower.  This is because the northern 
hemisphere of the secondary star is more easily seen and 
its southern hemisphere tends to be hidden from the observer 
if the orbital inclination is lower.

\subsection{Best-fitting parameter set for reproducing the observed orbital light curves}

We have performed orbital light curve simulations for 
the parameter sets with the orbital inclination 
$i$ = 30, 35, 40, 45, 50, and 55 deg and with the tilt angle 
$\theta$ = 0.5, 1.5, 2.5, 3.5, and 4.5 deg and four more sets 
for $\theta$ = 2.0~deg with the cases of $i$ = 45 and 50~deg 
and for $\theta$ = 3.0~deg with the cases of $i$ = 50 and 55~deg. 
The results are summarized in Table \ref{result-b}.  

\begin{table}
	\caption{Summary of the evaluation of our modeling results for the right figure of Figure \ref{orb8b} by the two criteria.  
	Here $\chi_{\nu}^2$ stands for reduced $\chi^2$.  }
	\vspace{2mm}
	\label{result-b}
	\centering
	\scalebox{0.9}{
	\begin{tabular}{|c|c|c|c|c|}
\hline
$i$ & $\theta$ & Semi-amplitude & $\chi^2$ & $\chi_{\nu}^2$ \\
\hline
30 & 0.5 & 0.005 & 150 & 0.95 \\ 
   & 1.5 & 0.014 & 132 & 0.84 \\ 
   & 2.5 & 0.023 & 164 & 1.04 \\ 
   & 3.5 & 0.032 & 244 & 1.54 \\ 
   & 4.5 & 0.041 & 375 & 2.38 \\ 
\hline
35 & 0.5 & 0.006 & 152 & 0.96 \\ 
   & 1.5 & 0.017 & 129 & 0.82 \\ 
   & 2.5 & 0.028 & 143 & 0.91 \\ 
   & 3.5 & 0.039 & 192 & 1.22 \\ 
   & 4.5 & 0.050 & 279 & 1.77 \\ 
\hline
40 & 0.5 & 0.007 & 154 & 0.98 \\ 
   & 1.5 & 0.020 & 129 & 0.81 \\ 
   & 2.5 & 0.033 & 131 & 0.83 \\ 
   & 3.5 & 0.047 & 160 & 1.01 \\ 
   & 4.5 & 0.060 & 217 & 1.37 \\ 
\hline
45 & 0.5 & 0.008 & 157 & 0.99 \\ 
   & 1.5 & 0.024 & 131 & 0.83 \\ 
   & 2.0 & 0.032 & 126 & 0.80 \\ 
   & 2.5 & 0.040 & 127 & 0.80 \\ 
   & 3.5 & 0.056 & 143 & 0.90 \\ 
   & 4.5 & 0.072 & 181 & 1.15 \\ 
\hline
50 & 0.5 & 0.010 & 159 & 1.00 \\ 
   & 1.5 & 0.029 & 133 & 0.84 \\ 
   & 2.0 & 0.038 & 126 & 0.80 \\ 
   & 2.5 & 0.048 & 124 & 0.79 \\ 
   & 3.0 & 0.057 & 125 & 0.79 \\ 
   & 3.5 & 0.067 & 131 & 0.83 \\ 
   & 4.5 & 0.086 & 155 & 0.98 \\ 
\hline
55 & 0.5 & 0.011 & 161 & 1.02 \\ 
   & 1.5 & 0.034 & 136 & 0.86 \\ 
   & 2.5 & 0.057 & 125 & 0.79 \\ 
   & 3.0 & 0.068 & 124 & 0.78 \\ 
   & 3.5 & 0.080 & 126 & 0.80 \\ 
   & 4.5 & 0.102 & 140 & 0.89 \\ 
\hline
\end{tabular}
}
\end{table}

As described in subsection 5.1, we use two criteria to 
judge which parameter set of simulations can best reproduce 
\textcolor{black}{the} observed orbital light curves for various super-orbital phases 
of the right figure of Figure \ref{orb8b}.  
The first criterion for the semi-amplitude of super-orbital 
light modulations basically measures the variation in 
the projected area of the tilted disk, and it reflects 
\textcolor{black}{
that the average light levels of individual orbital light 
curves vary in the super-orbital phase, as shown} 
in the right figure of Figure \ref{orb8b}.  
Its observational value is estimated to be 0.03 from 
Figure \ref{sup-b}.  

The second criterion is the goodness of fit between 
the observed light curves in the right figure of 
Figure \ref{orb8b} and simulated ones, represented by $\chi_{\nu}^2$.  
To calculate this, we first adjust the mean light level of 
each simulated orbital light curve to that of the observed one 
in the right figure of Figure \ref{orb8b} in order to examine 
purely the profile of the orbital light curves.  
We then calculate $\chi^2$ for each orbital light curve 
by the following equation: 
\begin{equation}
\chi^2 = \sum_{i = 1}^{N} \frac{(F_i - F(\varphi_i))^2}{\sigma_i^2}, 
\label{chi-squares}
\end{equation}
where $N$ is the number of the data per one orbital 
light curve, and in our case $N$ = 20, and 
$F_i$, $F(\varphi_i)$, and $\sigma_i$ represent 
the observed flux, the flux of the model light curve at 
a specific orbital phase $(\varphi_i)$, and the standard 
deviation of each observational data point, respectively.  
We calculate $\chi^2$ for one orbital light curve with 
a given super-orbital phase and sum them up 
for eight orbital light curves.  
The degree of freedom, $\nu$, is then 158 
because the number of total data points is 160 and 
the number of free parameters in our model is 2 for 
the orbital inclination and the tilt angle of the disk, 
respectively.  
We calculate the reduced $\chi^2$ (denoted as 
$\chi_{\nu}^2$) by dividing $\chi^2$ by $\nu$.  
The reduced $\chi^2$ is also given in 
Table \ref{result-b}.  
If it is around unity, the fit is supposed to be good.  
If it exceeds 1 greatly, the deviation of the model light 
curve from the observed light curve is too large.  
On the other hand, if it is extremely less than 1, 
it does not mean that the fit is better but it may simply 
mean that observational errors represented by $\sigma_i$ 
in equation (\ref{chi-squares}) are largely over-estimated.  
Since our minimum reduced chi-square is around 0.78 and 
close to 1, we consider that our modeling is good and 
that the smaller the chi-square is, the better the fit is.  

\begin{figure}[htb]
\begin{center}
\FigureFile(80mm, 50mm){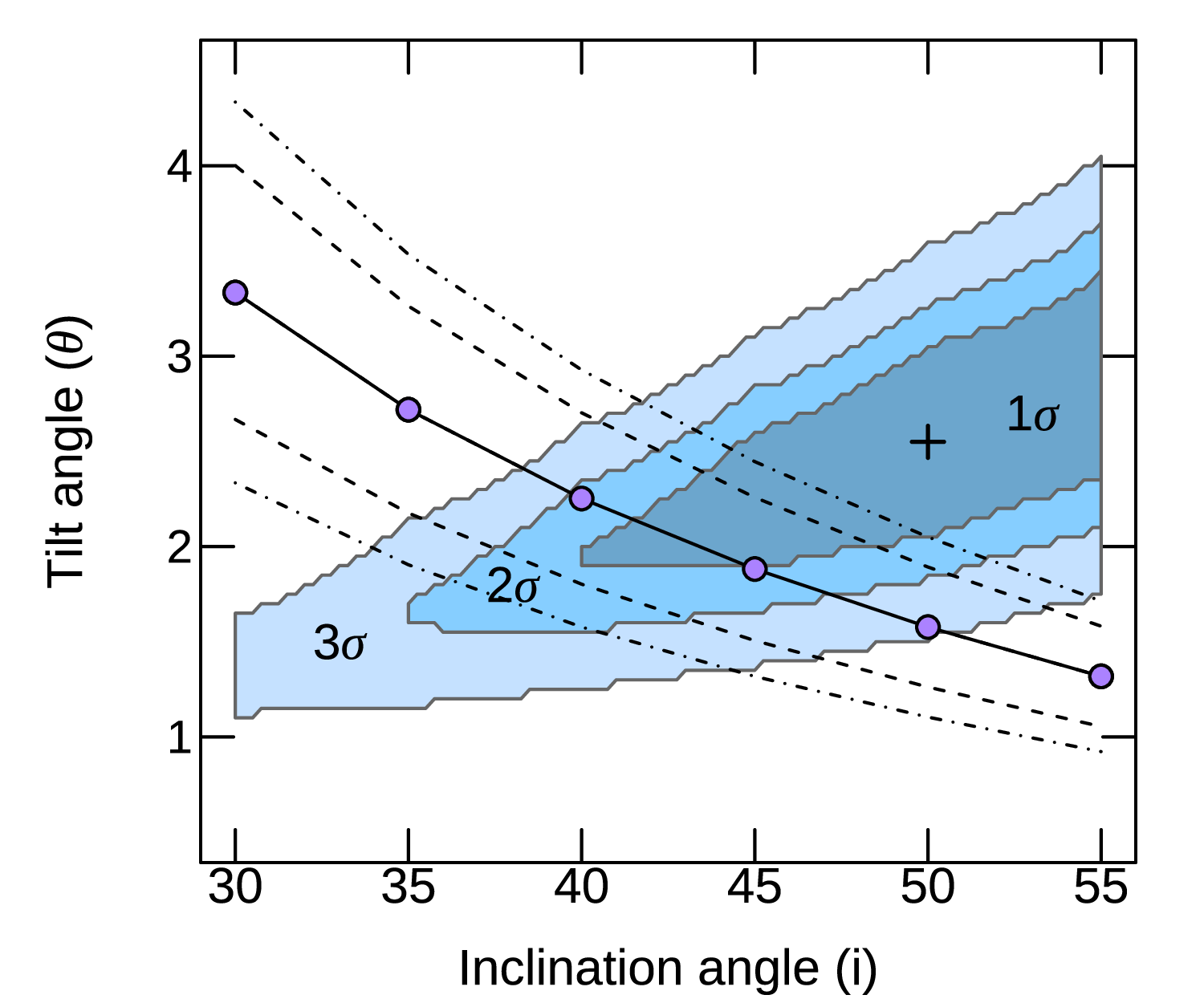}
\end{center}
\caption{Best-fitting parameters expected from the results in Table \ref{result-b}.  The dots stand for the best-fitting tilt angles for a given orbital inclination based on criterion 1, which are interpolated from the results of Table \ref{result-b}.  The cross is the best-fitting parameter set for criterion 2 yielding $\chi^2_{\rm min}$.  The error regions of $1\sigma$, $2\sigma$, and $3\sigma$ of the parameters, which are calculated on the basis of $\Delta \chi^2$ test, are drawn by blue contours.  If we allow a possible error of 20\% and 30\% for the semi-amplitude of super-orbital signals, the error ranges of the tilt angle for criterion 1 are represented as the areas surrounded by the dashed and dot-dashed lines.  
}
\label{inc-tilt}
\end{figure}

The results of our evaluation are shown in 
the ($i$, $\theta$) plane of Figure \ref{inc-tilt}.
We estimate the best-fitting tilt angle for criterion 1, 
which reproduces the observational semi-amplitude 
of super-orbital signals by linearly interpolating 
the results in Table \ref{result-b} for a given 
orbital inclination (the dots and the line 
connecting them).  
As discussed in subsection 5.3, some possible error 
is inevitable in determining the semi-amplitude of 
super-orbital signals.  
We must make some allowance for a possible range of 
the tilt angle for criterion 1.  
If we consider a possible error of 20\% and 30\% for 
the semi-amplitude with the value of 0.03,  
the possible ranges of the tilt angle are the areas 
surrounded by the dashed and dot-dashed lines in 
Figure \ref{inc-tilt}.  

As for criterion 2, we first set finer grids in 
the ($i$, $\theta$) plane and interpolate $\chi^2$ 
by using the results in Table \ref{result-b} at 
each grid point.  
We interpolate $\chi^2$ by a quadratic curve for 
each tilt angle and do linearly for each 
inclination angle.  
We then determine the best-fitting parameter set for 
$i$ and $\theta$ which minimizes $\chi^2$, and 
display it as the cross in 
Figure \ref{inc-tilt}.  
Here the minimum value of $\chi^2$ is denoted by  
$\chi^2_{\rm min}$. 
If we introduce $\Delta \chi^2$ 
defined as $\chi^2$ - $\chi^2_{\rm min}$, 
$\Delta \chi^2$ follows $\chi^2$ distribution with 
the degree of freedom of 2 \citep{DataReduction}.  
\textcolor{black}{
We plot by blue contours in Figure \ref{inc-tilt} 
the distributions of $\Delta \chi^2 \leq 2.30$, 
$\Delta \chi^2 \leq 4.61$, and $\Delta \chi^2 \leq 9.21$, 
which correspond to the 1$\sigma$, 2$\sigma$, and 3$\sigma$ 
error regions, respectively.}  

We see from Figure \ref{inc-tilt} that the two criteria 
give opposite dependence with respect to the orbital 
inclination and we find that the best-fitting parameter set 
is $i$ $\sim$ 45 deg and $\theta$ $\sim$ 2.0 deg to satisfy 
these two criteria.  
On the other hand, we have used much more information 
in evaluating criterion 2 than that by criterion 1.  
The weights of these two evaluations may not be equal.  
If we consider only the results of the profile fitting 
of light curves evaluated by criterion 2, 
the best-fitting parameter set is $i$ $\sim$ 50~deg and 
$\theta$ $\sim$ 2.5~deg.  
The acceptable ranges for the inclination angle and 
the tilt angle of the disk within 1$\sigma$ are around 
40--55~deg and 2.0--3.5~deg, respectively.
The case of low inclination angle, $\lesssim$35~deg, 
may be ruled out.

\section{Summary and discussion}

\subsection{Summary}

We have \textcolor{black}{first} reconstructed observational orbital 
light curves by subtracting negative superhump signals 
from the original data, as we did not do this in Paper I.  
The resulting light curves are found to be greatly modified 
from the original ones (see also Figure \ref{orb8b}). 
However, the main characteristics remain the same before 
and after this correction.  
We have performed the modeling of orbital light curves of 
KIC 9406652 by treating the retrogradely precessing tilted disk 
and the irradiated secondary star in order to reproduce 
the observed orbital signals in the brightening stage, 
whose profile varies as the super-orbital phase advances.
We have set the inclination angle of the binary system ($i$) 
and the tilt angle of the disk ($\theta$) as free parameters, 
and carried out simulations with various combinations of 
these parameter sets, and compared the results with 
the observational light curves displayed in 
the right figure of Figure \ref{orb8b}.  
We have found that the observed orbital curves are well 
reproduced by our modeling (e.g., Figure \ref{basic-i50-t1.5}).  
This confirms our interpretation of Paper I; 
observed orbital light curves are understood based on our model 
in which the secondary star is irradiated by the tilted 
and retrogradely precessing accretion disk, as the secondary star 
works as a reflecting mirror.  

We have explored the best-fitting parameter set of 
the orbital inclination, $i$, and the tilt angle, $\theta$,  
of the disk to reproduce the observed light curves.  
To do so, we have adopted two criteria: (1) the semi-amplitude 
of the super-orbital signal and (2) the goodness of fit 
of orbital light curve profiles between observed and 
calculated ones, represented by the $\chi^2$.
We have found that the best-fitting parameter set is something 
around $i \sim$~45~deg and $\theta \sim$~2.0~deg if we take 
into account these two criteria equally (see Figure \ref{inc-tilt}).  
However, the weights of these two criteria may not be necessarily 
equal as the profile fitting of criterion 2 uses 
more information than that of the semi-amplitude of 
criterion 1.  
If we use criterion 2 only, we have found that 
the best-fitting parameter set is $i \sim$~40--55~deg and 
$\theta \sim$2.0--3.5~deg.  The inclination angle lower 
than $\sim$35~deg would be ruled out from our results.  

In Paper I, we estimated the tilt angle of the disk 
only by the semi-amplitude of the super-orbital signal, 
which corresponds to our evaluation by criterion 1 
(see the line connecting the dots in Figure \ref{inc-tilt}).  
However, we figure out from our results that the tilt angle 
determined solely from this method could be underestimated.  
The estimation based on the light curve profile fitting gives 
a systematically higher value of the tilt angle for the orbital 
inclination of our interest (i.e., $i \sim 45-55$ deg) 
than that based on the semi-amplitude of the super-orbital 
signal.  

We may add here that we have performed the same modeling 
for a thicker disk having $H/r \sim 0.1$ at the disk 
outer edge and compared 
the results with those for a thinner disk of the standard case. 
Since the difference between the two is found to be so small, 
it may not be necessary to show the resulting figure.  
A tiny difference is that the irradiation effect 
is slightly larger in the case of the standard disk than 
that of the thicker one.  

\textcolor{black}{
We would like to discuss finally an interesting question 
raised by the referee; are there other negative-superhump stars 
with bigger amplitudes, which could in principle be used 
to carry out this kind of analysis ?  
In this respect, KIC 9406652, is unique in the sense 
that the complete orbital light curves are available 
for all phases of the super-orbital light modulations 
thanks to observations by the {\it Kepler} satellite. 
Besides, the irradiated component by the tilted accretion 
disk is clearly visible in its orbital light curves 
since its accretion disk keeps a bright state in most 
of the observational time and its inclination angle 
is neither too low nor too high in this star.  
Unfortunately, no other similar objects are known with 
such rich observational data available among CVs, 
as far as the authors know.  
On the other hand, there exists at least one star among 
X-ray binaries in which such rich data are available, 
which is HZ Her known as the optical counterpart of 
the X-ray source Her X-1 \citep{ger76hzher}.  
Her X-1/HZ Her is the famous X-ray binary with the orbital 
period of 1.7 d which shows 35-d X-ray ON-OFF cycle.  
This cycle is believed to be produced by occultation of 
the central X-ray source by the precessing tilted accretion disk.  
It is known that the optical orbital light curve of HZ Her varies 
in a highly systematic fashion over the 35-d cycle just 
in the same way as KIC 9406652.  
However, the source of irradiation to the secondary star is 
thought to be the central X-ray source in the case of HZ Her 
and the precessing tilted disk plays a role as a shadowing 
object for the irradiated secondary star.  
In this respect, the situation is different between KIC 9406652 
and HZ Her.  
Also, the orbital inclination of HZ Her is higher than 80~deg 
\citep{lea14hzher}.  
We hope that future observations could provide 
such rich data for the negative-superhump objects among CVs 
and the present type of analysis could have a promising return.  
For example, some of the bright CVs such as V1193 Ori and 
V751 Cyg listed in Table 5 of \citet{arm13aqmenimeri} have 
been observed by the Transiting Exoplanet Survey Satellite 
({\it TESS}) known as the successor of the {\it Kepler} satellite, 
and hence, this kind of analysis would be applicable to 
these objects once abundant data is collected.  
}

\subsection{Consistency in the estimation of the inclination angle between photometric and spectroscopic approaches}

As discussed in subsection 4.1, the orbital inclination 
of this system is estimated from the spectroscopic 
considerations and it is around $i$ = 50~deg 
\citep{gie13j1922}.  
If uncertainty in the spectroscopic observations is 
taken into account, its possible range \textcolor{black}{is} 45--60~deg.  
Our estimate of the orbital inclination around 45~deg 
based solely on the photometric considerations is thus 
consistent with that from the spectroscopic considerations.  

Strictly speaking, the best-fitting inclination 
angle, 45~deg, via our photometric approach is slightly 
lower than that of 50~deg derived by \citet{gie13j1922} 
via optical spectroscopy. 
If the difference between these two values matters 
and if the blame lies on the photometric side,  
we may here explore possible causes why our modeling yields 
a lower inclination angle.  
One possibility is that we were not able to sufficiently 
subtract the effects of negative superhumps in section 2.  
If the effects of negative superhumps had still remained in 
the observational light curves shown in the right panel 
of Figure \ref{orb8b}, and if we could remove these effects 
completely, the phase shift of the light maximum 
and the amplitude variation in orbital light curves would become 
smaller with the advance of super-orbital phases.  
The blue contour of Figure \ref{inc-tilt} would then be 
shifted towards smaller tilt angles in such a case. 
As a result, the best-fitting inclination angle would approach 
towards 50~deg. 
Another possibility \textcolor{black}{is} that the semi-amplitude of 
super-orbital signals could be underestimated.  
This case would also give a larger inclination angle.

\subsection{Disappearance of the orbital hump during brightening}

In this study, we have treated mainly the irradiated 
secondary star in the orbital light curve modeling and 
we have not considered the orbital hump which is supposed 
to be produced by the collision of the gas stream 
at the disk rim.  
In fact, the orbital signal during the brightening stage is 
dominated by the effect of irradiation of the secondary star
and no signal of the orbital hump is visible 
since the averaged orbital light curve exhibited in 
Figure \ref{orb-b} peaks at the orbital phase 0.5 and 
is symmetric with respect to that phase.

On the other hand, we detected the orbital hump peaked 
at phase 0.8--0.9 in the orbital signal during 
quasi-standstills (see the middle panel of Fig.~8 in 
Paper I).  
If the luminosity of the bright spot and that of 
the disk outer edge are the same as those 
in quasi-standstills, we \textcolor{black}{see} an orbital hump with 
the same amplitude in the flux scale during brightening 
as that in quasi-standstills.  
However, the orbital hump seems to have completely disappeared  
when the system brightness increases only by $\sim$1.6 times.  
That looks rather strange.  

In ordinary U Gem-type stars, the orbital hump 
(which is a dominant feature in the orbital light curve 
in quiescence) disappears during outburst, because 
the flux ratio of the bright spot with respect to 
the disk outer edge drastically increases 
in comparison with that in quiescence.
One possibility is that the outer part of the disk at 
quasi-standstills in KIC 9406652 could be in the cool state 
so that the bight spot is conspicuous.  
In fact, we proposed the model for the light variation of 
IW And-type stars including this object in 
\citet{kim20tiltdiskmodel}, in which only the outer part of 
the disk is in the cool state during quasi-standstills.  
However, this was not supported by the observed 
frequency variation of negative superhumps, which suggests 
that the disk gradually expands during quasi-standstills 
(see Paper I).  
If the outer disk is cool, the disk gradually shrinks 
due to the inflow of material having small angular momentum 
to the disk outer edge.  
The observation of negative superhumps rather implies that 
the outer disk is in the hot state during quasi-standstills.  
We have no good idea for explaining the disappearance of 
orbital humps during brightening and it may remain one 
of the future problems.

\section*{Acknowledgements}

M.~Kimura acknowledges support by the Special Postdoctoral 
Researchers Program at RIKEN.  
The original numerical code that we revised in this paper 
was provided by Izumi Hachisu.  
This work was financially supported by Japan Society for 
the Promotion of Science Grants-in-Aid for Scientific 
Research (KAKENHI) Grant Number JP20K22374 (MK).  
We thank the anonymous referee for helpful comments.


\newcommand{\noop}[1]{}


\end{document}